\documentclass[aps,prc,showpacs,floatfix,superscriptaddress,twocolumn]{revtex4}

\preprint{JLAB-THY-11-...}

\usepackage{amsmath}
\usepackage{bm}
\usepackage{graphicx}

\begin{document}

\title{New data strengthen the connection between Short Range Correlations and the EMC effect}

\author{O.~Hen}
\affiliation{Tel Aviv University, Tel Aviv 69978, Israel}

\author{E.~Piasetzky}
\affiliation{Tel Aviv University, Tel Aviv 69978, Israel}

\author{L. B.~Weinstein}
\affiliation{Old Dominion University, Norfolk, Virginia 23529, USA}

\date{\today}

\begin{abstract}
Recently published measurements of the two nucleon short range correlation ($NN$-SRC) scaling factors, $a_2(A/d)$, strengthen the previously observed correlation between the magnitude of the EMC effect measured in electron deep inelastic scattering at $0.35\le x_B\le 0.7$ and the SRC scaling factor measured at $x_B \ge 1$.  The new results have improved precision and include previously unmeasured nuclei. The measurements of $a_2(A/d)$ for $^9$Be and $^{197}$Au agree with published predictions based on the EMC-SRC correlation.  
This paper examines the effects of the new data and of different corrections to the data on the slope and quality of the EMC-SRC correlation, the size of the extracted deuteron IMC effect, and the free neutron structure function. The results show that the linear EMC-SRC correlation is robust and that the slope of the correlation is insensitive to most combinations of corrections examined in this work. This strengthens the interpretation that both $NN$-SRC and the EMC effect are related to high momentum nucleons in the nucleus.
\end{abstract}

\pacs{25.30.Fj,13.60.Hb,21.30.-x}

\maketitle

%%%%%%%%%%%%%%%%%%%%%%%%%%%%%%%%%%%%%%%%%%%%%%%%%%%%%%%%%%%%%%%%%%%%%%%%%

\section{Introduction}
The per-nucleon lepton deep inelastic scattering (DIS) cross sections of heavy nuclei are less than those of deuterium at moderate to large four-momentum transfer, $Q^2\ge 2$ (GeV/c)$^2$, and $0.35 \le x_B \le 0.7$ ($x_B=Q^2/2m\nu$, where $\nu$ is the energy transfer and $m$ is the proton mass). This ``EMC effect'' was discovered in 1982 in the cross section ratios of iron to deuterium \cite{Aubert83} and confirmed by many measurements on a range of nuclei \cite{Ashman88,Gomez94,Arneodo88,Arneodo90,Allasia90,Seely09}. The latest data \cite{Seely09} show that for light nuclei the EMC effect does not increase monotonically with increasing average nuclear density.  While there is no generally accepted explanation of the EMC effect, proposed explanations generally need to include both nuclear structure effects (momentum distributions and binding energy) and modification of the bound nucleon structure.

A recent paper showed a strong correlation between the magnitude of the EMC effect and the Short Range Correlations (SRC) scaling factor \cite{weinstein11}.  Because the per-nucleon cross section ratios of nuclei to deuterium for $0.35 \le x_B \le 0.7$ decrease approximately linearly with $x_B$, in this range of $x_B$ the EMC effect can be quantified by the slope of this ratio, $dR_{EMC}/dx_B$ \cite{Seely09}. The SRC scaling factor, $a_2(A/d)$, equals the ratio of the per-nucleon inclusive electron scattering cross section for nucleus $A$ to deuterium at $Q^2 > 1.4$ (GeV/c)$^2$ and $1.5 \le x_B < 2$.  In this range of $x_B$, the cross section ratio is constant \cite{day87,frankfurt93,egiyan03,egiyan06}.  The constancy of the ratio in this range of $x_B$ is attributed to high-momentum components of the nuclear wave function.  These high momentum components have been shown to be almost entirely due to central and tensor nucleon-nucleon short range correlations \cite{tang03,piasetzky06,shneor07,subedi08,baghdasaryan10}.

This correlation between the magnitude of the EMC effect measured at $0.35 \le x_B \le 0.7$ and the SRC scale factor measured at $1.5 \le x_B < 2$ was used to phenomenologically determine the ratio of the DIS cross section for a proton and neutron bound in deuterium to the DIS cross section for free (unbound) $pn$ pair and thus to determine the free neutron cross section for $0.35 \le x_B \le 0.7$.  The free neutron cross section was then used to determine the ratio of the neutron to proton structure function, $F_2^n/F_2^p$ and hence the ratio of $d/u$ in this range of $x_B$.

Recently, high precision measurements of the per nucleon inclusive electron scattering cross section ratio for different nuclei relative to deuterium at $Q^2 \sim 2.7$ (GeV/c)$^2$ and $1>x_B>2$ were published~\cite{fomin12}, covering more nuclei at greater precision than previous measurements.
These ratios also show scaling behavior for $x_B>1.5$.   This new data allows us to reexamine the observed linear correlation between the strength of the EMC effect and the SRC scaling factor~\cite{weinstein11}. 

The analysis of the new data also includes various corrections to the measured cross section ratios that were not included in previous analyses ~\cite{egiyan03,egiyan06}. 

In this paper we examine the consistency of the old and new data and the effects of different corrections to the cross section ratios and therefore on the slope of the EMC-SRC correlation.  We also examine the effects of these on the ratio of the bound to free $pn$ DIS cross sections and on the free neutron structure function~\cite{hen11}.

%%%%%%%%%%%
\begin{table*}[t]
\begin{center}
\caption{A comparison of SRC scaling factors, $a_2(A/d)$, extracted from different data sets with different corrections.  Column 2 shows the scaling factors from Egiyan {\it et al} \cite{egiyan06}.  Column 3 shows the prediction of Ref.~\cite{weinstein11} based on the EMC data of Refs.~\cite{Gomez94,Seely09}. Columns 4 through 6 show the data of Fomin {\it et al.} \cite{fomin12} with different corrections.  Column 4 shows the data with the same corrections used in Egiyan {\it et al.}, column 5 shows the data as published, and column 6 shows the data excluding their correction for the center of mass motion of the SRC-pair.  Column 7 shows the results from SLAC \cite{frankfurt93}. Column 8 shows the slopes of the EMC effect from Refs.~\cite{Gomez94,Seely09} as cited in \cite{weinstein11}.  See the text for more details. \\
$^*$The $^3$He SRC scaling factor in column 2 from Ref.~\cite{egiyan06} was determined primarily from the calculated ratio of the $^3$He and $d$ momentum distribution above the scaling threshold ($p_{thresh}=0.275\pm0.025$ GeV/c).  \\
$^{**}$The SLAC ratios~\cite{frankfurt93} used cross sections from different experiments at different kinematics.  They interpolated the deuterium cross sections to the kinematics of the cross sections measured for heavier nuclei and have larger uncertainties than the later measurements.  They are included here for completeness.
 }
\begin{tabular}{|l|l| l | l | l | l | l | l |}
    \hline
          & Egiyan {\it et al.} & EMC-SRC                    & Fomin {\it et al.}~\cite{fomin12} & Fomin {\it et al.}  & Fomin {\it et al.}~\cite{fomin12}  &    SLAC & EMC Slope~\cite{weinstein11} \\
                & \cite{egiyan06} & Prediction~\cite{weinstein11}  & [Analysis as in           &     \cite{fomin12} & [excluding the CM & \cite{frankfurt93}$^{**}$ &  $dR_{EMC}/dx$   \\
   Nucleus &                      &                            & Ref.~\cite{egiyan06}]       &                            &   motion correction]  &  &          \\
    \hline
    column $\#$      & \multicolumn{1}{c|}{2}     & \multicolumn{1}{c|}{3}     & \multicolumn{1}{c|}{4}    & \multicolumn{1}{c|}{5}     & \multicolumn{1}{c|}{6} & \multicolumn{1}{c|}{7} & \multicolumn{1}{c|}{8} \\
    \hline
    $^3$He     & $1.97 \pm 0.10^*$             &                            & $1.87 \pm 0.06$           & $1.93 \pm 0.10$          & $2.13 \pm 0.04$ & $1.7\pm0.3$    & $-0.070\pm0.029$    \\
    $^4$He      & $3.80 \pm 0.34$             &                            & $3.64 \pm 0.07$           & $3.02 \pm 0.17$          & $3.60 \pm 0.10$ & $3.3\pm0.5$   & $-0.197\pm0.026$    \\
    $^9$Be     &                            & $4.08 \pm 0.60$            & $4.15 \pm 0.09$           & $3.37 \pm 0.17$          & $3.91 \pm 0.12$ &  & $-0.243\pm0.023$    \\
    $^{12}$C     & $4.75 \pm 0.41$             &                        & $4.81 \pm 0.10$           & $4.00 \pm 0.24$          & $4.75 \pm 0.16$ & $5.0\pm0.5$  & $-0.292\pm0.023$    \\
    $^{56}$Fe($^{63}$Cu) & $5.58 \pm 0.45$        &                            & $5.29 \pm 0.12$           & $4.33 \pm 0.28$          & $5.21 \pm 0.20$ & $5.2\pm0.9$ & $-0.388\pm0.032$    \\
    $^{197}$Au     &                             & $6.19 \pm 0.65$            & $5.29 \pm 0.16$           & $4.26 \pm 0.29$          & $5.16 \pm 0.22$ & $4.8\pm0.7$ & $-0.409\pm0.039$    \\
    \hline
    \hline
    EMC-SRC slope $a$      & $0.079 \pm 0.006$     &      & $0.082 \pm 0.004$         & $0.106 \pm 0.006$        & $0.084 \pm 0.004$  &        &      \\
    $\frac{\sigma(n+p)}{\sigma_d}|_{x_B=0.7}$ & $1.032 \pm 0.004$     &      & $1.033 \pm 0.004$       & $1.043 \pm 0.005$        & $1.034 \pm 0.004$   & &      \\
    $\chi^2/ndf$                  & $0.7688 / 3$          &      & $4.742 / 5$               & $4.078 / 5$              & $4.895 / 5$                &     &      \\
    \hline
\end{tabular}
\end{center}
\label{tab:results}
\end{table*}

\section{The New Data}
New measurements by Fomin {\it et al.}~\cite{fomin12} of the SRC scaling factor $a_2(A/d)$ have about four times smaller uncertainties than previous ones by Egiyan {\it et al.}~\cite{egiyan03,egiyan06}.  They also include two nuclei, $^9$Be and $^{197}$Au for which the SRC scaling factors were previously predicted based on their measured EMC effect~\cite{Gomez94,Seely09} and the linear EMC-SRC correlation \cite{weinstein11}. $^9$Be is of particular interest due to the anomalous density dependence of its EMC effect (its EMC effect is larger than that of $^4$He although its average density is much smaller)~\cite{Seely09}. It therefore presents a challenging test for the prediction made in~\cite{weinstein11} and for the validity of the EMC-SRC correlation in general.

The different measurements have different corrections applied to their results.  Both sets of measurements applied radiative corrections to their measured cross section ratios.  Egiyan {\it et al.} \cite{egiyan03,egiyan06} also applied isoscalar corrections to correct for differences in the per nucleon cross section ratio for asymmetric nuclei due to the difference between the elementary electron-proton and electron-neutron cross sections.  Fomin {\it et al.} ~\cite{fomin12} did not apply the isoscalar correction but did apply corrections for the nuclear coulomb field, inelastic contributions, and SRC-pair center of mass motion. Inspired by results of exclusive $^{12}$C$(p,ppn)$ and $^{12}$C$(e,e'pN)$ measurements, which showed that $NN$-SRC pairs are dominated by neutron-proton pairs ($\sim 18$ times more neutron-proton then proton-proton pairs were observed)~\cite{tang03,piasetzky06,shneor07,subedi08,baghdasaryan10}, Fomin {\it et al.} assumed that at $x_B>1.4$, electrons scatter mainly off neutron-proton pairs and therefore isoscalar corrections are unnecessary. The largest correction made by Fomin {\it et al.} is a correction for enhancement of the cross section ratio (and therefore of the SRC scaling factors) due to the SRC-pair center of mass (c.m.) motion for $A>2$. The c.m. correction is defined as the ratio of the convolution of the pair c.m. motion and deuteron momentum distributions to the deuteron momentum distribution. This ratio was calculated in \cite{fomin12} for $^{56}$Fe using the SRC-pair momentum distributions of Ciofi degli Atti and Simula \cite{cda96}.  It was then scaled to other nuclei based on the $A$-dependence of the pair motion. Due to uncertainties in the calculation, including its $x_B$ and $A$ dependence, they applied an uncertainty equal to  $30-50\%$ of the calculated correction.

Table~\ref{tab:results} lists the per nucleon cross section ratios for all nuclei measured by Fomin {\it et al.} The second column shows the ratios measured by Egiyan {\it et al.} that were used in the original EMC-SRC analysis \cite{weinstein11}. Fomin {\it et al.} measured $^{63}$Cu, which was not measured by Egiyan  {\it et al.}; we assume the SRC scaling factor of $^{63}$Cu to be the same as that of $^{56}$Fe. The values of $^9$Be and $^{197}$Au in the third column are those predicted by Ref.~\cite{weinstein11} based on their measured EMC effect and the linear EMC-SRC correlation. The fourth column shows the Fomin {\it et al.} results, analyzed in the same manner as the Egiyan data (i.e., including radiative and isoscalar corrections only). The fifth column shows the Fomin {\it et al.} results as published (i.e., including inelastic, radiative, coulomb, and center of mass motion corrections). The sixth column shows the as published Fomin {\it et al.} results with  the center of mass motion correction removed (i.e., including inelastic, radiative, and coulomb corrections). Comparing the second and fourth columns, one can see that the measured values of $a_2(A/d)$ from the two measurements agree within uncertainty when analyzed with the same corrections (radiative and isoscalar corrections only).  Applying the radiative, coulomb field and inelastic (but not the isoscalar) corrections changes the measured scale factors by about 10\%.  Applying the SRC-pair center of mass motion correction decreases the ratios by 10\% to 20\%.  The last column of Table~\ref{tab:results} shows the magnitude of the EMC effect for the different nuclei as measured by \cite{Gomez94,Seely09} and averaged by \cite{weinstein11}.  
%Following \cite{Seely09}, the magnitude of the EMC effect is determined by the slope of the per-nucleon DIS cross section ratio of nucleus $A$ to deuterium for $0.35\le x_B\le 0.7$.

\section{The EMC-SRC correlation}
\begin{figure}[htb]
\includegraphics[height=8cm,width=8.5cm]{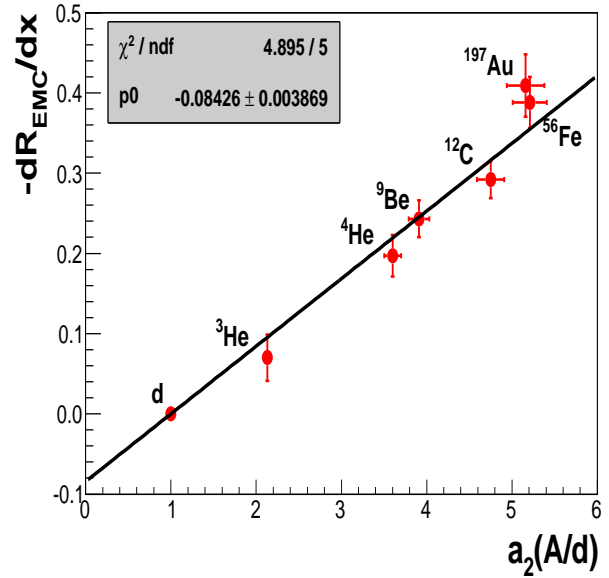}
\caption{The slope of the EMC effect for $0.35 \le x_B \le 0.7$ plotted versus $a_2(A/d)$, the SRC scaling factor (the relative amount of 2N-SRC pairs) in a variety of nuclei.  The uncertainties include both statistical and systematic uncertainties added in quadrature.  The values of $a_2(A/d)$ are taken from Fomin {\it et al.} \cite{fomin12} as published except for the SRC-pair center of mass motion corrections.  The fit parameter, $a=-0.084\pm0.004$, is the intercept of the line and also the negative of the slope of the line.}
\label{fig:EMC_SRC_new}
\end{figure}

The quality of the correlation between the magnitude of the EMC effect and the newly measured SRC scaling factors, $a_2(A/d)$, is shown in Fig.~\ref{fig:EMC_SRC_new}.  Due to the large uncertainties of the SRC-pair center-of-mass motion correction, Fig.~\ref{fig:EMC_SRC_new} shows the data of Fomin {\it et al.} as published but without that correction.  Fig.~\ref{fig:EMC_SRC_new} also shows the results of a one-parameter fit to the EMC slopes as a function of the SRC scaling factors.   Because the point for the deuteron is fixed at $dR_{EMC}/dx=0$ and $a_2(A/d)=1$, the fitted slope is also the negative of the intercept of the line.

In order to test the robustness of the EMC-SRC correlation, we made a series of one-parameter linear fits to the EMC slopes (Table \ref{tab:results} column 8) as a function of the different SRC scaling factors shown in Table~\ref{tab:results}.  The $\chi^2$ per degree of freedom for each of these fits was approximately one, indicating an excellent fit.  In addition, the values of $a_2(A/d)$ predicted for $^9$Be and $^{197}$Au by Ref.~\cite{weinstein11} agree within uncertainties with the new values measured by Fomin {\it et al.} with the radiative and isoscalar corrections from \cite{egiyan06}.

%%%%%%%%%%%%%%
%%%%%%%%%%%%%%

Following Ref.~\cite{weinstein11}, the value $a_2(A/d)=0$ corresponds to the limit of free nucleons with no SRC.  If we extrapolate the linear fit to this point, this should give us the EMC ratio for a free (unbound) $pn$ pair to the deuteron, the so-called In-Medium Correction (IMC) effect.  The IMC effect then equals the negative of the fitted EMC-SRC slope.  This value ranges from $\vert dR_{IMC}/dx\vert =0.079\pm0.006$ to $0.084\pm0.004$ for the different data sets with the different corrections (excluding the cm motion correction).  If we include the SRC-pair center of mass motion correction, then the linear fit is still excellent. However the slope and hence the intercept increases by about 20\% to $0.106\pm0.006$.

Since the EMC effect is linear for $0.3 \le x_B \le 0.7$, we have (also following \cite{weinstein11}),
\[
\frac{\sigma_d}{\sigma_p+\sigma_n}=1-a(x_B-b)
\]
where $\sigma_d$ and $\sigma_p$ are the measured DIS deuteron and proton cross sections, $\sigma_n$ is the unmeasured free neutron cross section, $a=\vert dR_{IMC}/dx\vert\approx 0.08$ and $b=0.31\pm0.04$ is the average value of $x_B$ where the EMC effect is unity (i.e., where the per-nucleon cross sections are equal).  Evaluating this at $x_B=0.7$ gives the ratio of the free $pn$ cross section to the bound $pn$ (deuteron) cross section which ranges from $1.032\pm0.004$ to $1.034\pm0.004$ for the different data sets and corrections (again excluding the cm motion correction).  If we include the cm motion correction, then this ratio changes to $1.043\pm0.005$.

The agreement of the slope of the EMC-SRC correlation, and therefore of the deuteron IMC effect at $x_B=0.7$, among all combinations of data sets and corrections is a clear indication of the robustness of the EMC-SRC correlation. This also indicates that the deuteron IMC effect and the free neutron structure function extracted in \cite{weinstein11} and used in \cite{hen11} do not change due to the new data and/or analysis. If the center of mass motion correction is included, then the linearity of the EMC-SRC relation improves slightly and the deuteron IMC effect increases by about 20\% to $dR_{IMC}/dx = 0.106 \pm 0.006$.

\section{Conclusions and Outlook}
New higher precision data \cite{fomin12} strengthens the phenomenological correlation between the strength of the EMC effect and the relative amount of SRC-correlated $NN$ pairs in a nucleus~\cite{weinstein11}. The new measurements are consistent with the SRC scaling factors for $^9$Be and $^{197}$Au that were predicted based on this EMC-SRC correlation. Different corrections for the SRC cross section ratio were examined and all were shown to be consistent with a linear correlation between the strength of the EMC effect and the relative amount of SRC correlated $NN$ pairs in nuclei. The linearity of the EMC-SRC correlation, regardless of the exact corrections considered, is a clear indication of the robustness of the EMC-SRC correlation.

This strengthens the speculation presented in \cite{weinstein11} that both the EMC effect and $NN$-SRC  originate from high momentum nucleons in the nucleus.

More data is required to further map out and understand this correlation.  Several experiments approved to run as part of the 12 GeV program at Jefferson Lab will measure both the SRC scaling factors and the EMC effect at high precision over a wide range of light and heavy nuclei \cite{arringtonexpt06,arringtonexpt10,solvignonexpt11}.   Another experiment~\cite{Emcsrcexpt11} will search for medium modification of the structure function of deeply bound, high momentum, nucleons. This will be done by performing DIS scattering off high momentum nucleons in deuterium and tagging the partner (high momentum) recoil nucleon. The results of this experiment will allow to compare the structure function of free and bound nucleons and gain insight on the connection of the EMC effect to high momentum nucleons in the nucleus.

%%%%%%%%%%%%%%%%%%%%%%%%%%%%%%%%%%%%%%%%%%%%%%%%%%%%%%%%%%%%%%%%%%%%%%%%%
%\bibliography{eep,emc}

\end{document}